\documentclass[11pt,letterpaper]{article}
\pdfoutput=1
\usepackage{jcappub}

\usepackage{appendix}
\usepackage{array}
\newcolumntype{x}[1]{>{\centering\arraybackslash}p{#1}}

\usepackage{epsfig}
\usepackage{graphicx}
\usepackage{dcolumn}
\usepackage{amsmath}
\usepackage{enumerate}

\usepackage{color}
\usepackage{ulem}
\hypersetup{colorlinks=false}

\title{Halo-independent analysis of direct detection data for light WIMPs}

\author[a]{Eugenio Del Nobile,}
\author[a]{Graciela B. Gelmini,}
\author[b]{Paolo Gondolo,}
\author[a]{and Ji-Haeng Huh}

\affiliation[a]{Department of Physics and Astronomy, UCLA,\\
475 Portola Plaza, Los Angeles, CA 90095, USA}
\affiliation[b]{Department of Physics and Astronomy, University of Utah,\\
115 South 1400 East \#201, Salt Lake City, UT 84112, USA}

\emailAdd{delnobile@physics.ucla.edu}
\emailAdd{gelmini@physics.ucla.edu}
\emailAdd{paolo@physics.utah.edu}
\emailAdd{jhhuh@physics.ucla.edu}

\abstract{We present a halo-independent analysis of direct detection data on ``light WIMPs," i.e.\ weakly interacting massive particles with mass close to or below 10 GeV/$c^2$. We include new results from silicon CDMS detectors (bounds and excess events), the latest CoGeNT acceptances, and recent measurements of low sodium quenching factors in NaI crystals. We focus on light WIMPs with spin-independent isospin-conserving and isospin-violating interactions with nucleons. For these dark matter candidates we find that a low quenching factor would make the DAMA modulation incompatible with a reasonable escape velocity for the dark matter halo, and that the tension among experimental data tightens in both the isospin-conserving and isospin-violating scenarios. We also find that a new although milder tension appears between the CoGeNT and DAMA annual modulations on one side and the silicon excess events on the other, in that it seems difficult to interpret them as the modulated and unmodulated aspects of the same WIMP dark matter signal.}

\keywords{dark matter theory, dark matter experiments}

\begin{document}

\maketitle

\section{Introduction}
The nature of dark matter is one of the  fundamental  unsolved problems of physics and cosmology.  Among the dark matter candidates most actively searched for are weakly interacting massive particles (WIMPs), which are  particles with weakly interacting cross sections
and masses in the GeV/$c^2$ --10~TeV/$c^2$ range.  Of particular interest are WIMPs with mass
around 1-10 GeV/$c^2$, the so-called ``light WIMPs.''

Many theoretically-viable light WIMPs with mass $\lesssim 10$ GeV/$c^2$ have been suggested since the late 1970s, ranging from heavy massive neutrinos to supersymmetric particles (for early papers, see e.g.~\cite{lightWIMPtheory}). Light WIMPs with spin-independent interactions were first singled out in 2004~\cite{Gelmini:2004gm} as a viable WIMP interpretation of the DAMA annual modulation~\cite{Bernabei:2003za}
 compatible with all direct searches at the time. The interest in light WIMPs intensified after the first  DAMA/LIBRA results of 2008~\cite{Bernabei:2008yi}, and even more in recent years due to other potential dark matter hints in the same region.

Four direct dark matter search experiments (DAMA~\cite{Bernabei:2010mq}, CoGeNT~\cite{Aalseth:2010vx, Aalseth:2011wp, Aalseth:2012if},  CRESST-II~ \cite{Angloher:2011uu}, and now CDMS-II~\cite{Agnese:2013rvf})  have data  that may be interpreted as  signals from light WIMPs. DAMA~\cite{Bernabei:2010mq} and CoGeNT~\cite{Aalseth:2011wp} report annual modulations in their event rates compatible with those expected for a dark matter signal~\cite{Drukier:1986tm}. CoGeNT~\cite{Aalseth:2010vx,  Aalseth:2012if}, CRESST-II~ \cite{Angloher:2011uu}, and very recently, CDMS-II~\cite{Agnese:2013rvf} observe an excess of events above their expected backgrounds; the excess may be interpreted as due to dark matter WIMPs. In particular, the CDMS-II collaboration, in their latest analysis of their silicon detectors~\cite{Agnese:2013rvf}, has very recently reported three events in their WIMP signal region against an expected background of $0.7$ events. Data were taken between 2007 and 2008 with 8 silicon detectors, in the nuclear recoil energy range $7$ to $100$ keV, for a total exposure of 140.2 kg-day prior to the application of selection cuts. 

On the other hand,  upper limits have been placed on dark matter WIMPs by other direct detection experiments. The most stringent limits on the average (unmodulated) rate for light WIMPs come from the XENON10~\cite{Angle:2011th}, XENON100~\cite{Aprile:2011hi, Aprile:2012nq} and CDMS-II-Ge experiments~\cite{Ahmed:2010wy} with the addition of SIMPLE~\cite{Felizardo:2011uw} in the isospin-violating case. The amplitude of an annually modulated signal is directly constrained by CDMS-II-Ge~\cite{Ahmed:2012vq}.

In the following we compare the different experimental results in a way independent of the dark halo model \cite{Fox:2010bz, Frandsen:2011gi, Gondolo:2012rs, Frandsen:2013cna, DelNobile:2013cva, HerreroGarcia:2011aa, HerreroGarcia:2012fu, Bozorgnia:2013hsa}) for dark matter candidates with the standard spin-independent WIMP-nucleus interaction, both with isospin-conserving and isospin-violating couplings.

\section{The halo-independent analysis}

We follow the specific method presented in Ref.~\cite{Gondolo:2012rs}, which takes into account form factors, experimental  energy resolutions, acceptances, and efficiencies with arbitrary energy dependence.

The differential recoil rate per unit detector mass,  typically  in units of
counts/kg/day/keV, for the scattering of WIMPs of mass $m$ off 
nuclei of mass number $A$, atomic number $Z$, and mass $m_{A,Z}$ is
\begin{equation} 
  \frac{dR_{A,Z}}{dE}  = \frac{\sigma_{A,Z}(E)}{2 m \mu_{A,Z}^2}\, \rho\, \eta(v_{\rm {min}},t) ,
  \label{dRdE}
\end{equation}
where $E$ 
is the nucleus recoil energy,  $\rho$ is the local WIMP density, $\mu_{A,Z} = m_{A,Z} \, m / (m_{A,Z} + m)$ is the WIMP-nucleus reduced mass,
 $\sigma_{A,Z}(E)$ is (a multiple of)
the WIMP-nucleus differential cross-section $d\sigma_{A,Z}/dE= \sigma_{A,Z}(E)~m_{A,Z}/  2\mu_{A,Z}^2 v^2$, and
\begin{align}
\eta(v_{\rm min},t) = \int_{|{\bf v}|>v_{\rm min}} \frac{f({\bf v},t)}{v} d^3 v
\end{align}
is a velocity integral carrying the only dependence on the (time-dependent) distribution $f({\bf v},t)$ of WIMP
velocities ${\bf v}$ relative to the detector. Here
\begin{equation}
 v_{\rm {min}} = \sqrt{\frac{m_{A,Z} E}{2\mu_{A,Z}^2}}
  \label{vmin}
\end{equation}
is the minimum WIMP speed that can result in a recoil energy $E$ in an elastic scattering with the $A,Z$ nucleus.
Due to the revolution of the Earth around the Sun, the $\eta$ function has an annual modulation
generally well approximated by the first  terms of a harmonic series
\begin{equation}
  \eta(v_{\rm {min}},t) = \eta_0(v_{\rm {min}}) + \eta_1(v_{\rm {min}}) \cos\!\left[ \omega (t-t_0) \right],
   \label{eta} 
\end{equation}
where $\omega = 2\pi$/yr and $t_0$ is the time of maximum signal.

For spin-independent interactions (SI),  the WIMP-nucleus cross-section can be written in terms of the effective WIMP-neutron and WIMP-proton coupling constants $f_n$ and $f_p$ as
\begin{equation} 
  \sigma^{SI}_{A,Z}(E) =  \sigma_p \frac{ \mu_{A,Z}^2}{ \mu_p^2} [Z+ (A-Z)(f_n/f_p)]^2\,   \,F_{A,Z}^2(E) \, ,
  \label{sigma}
\end{equation}
where $\sigma_p$ is the WIMP-proton cross-section and 
$F_{A,Z}^2(E)$ is a nuclear form factor, which we take to be a Helm form factor~\cite{Helm:1956zz} normalized to $F_{A,Z}(0) = 1$.
In most  models the couplings are isospin-conserving, $f_n = f_p$.
Isospin-violating couplings $f_n \ne f_p$ have been considered as a possibility to weaken the upper bounds  obtained with heavier target elements, which being richer in neutrons than lighter elements, have their couplings to WIMPs suppressed for $f_n/f_p \simeq -0.7~$\cite{Kurylov:2003ra}. 

It is worth noticing that the function
\begin{equation} 
   \tilde\eta(v_{\rm min}) = \frac{\sigma_p\rho}{m} \,  \eta(v_{\rm min})
     \label{tilde-eta}
\end{equation}
is common to all direct detection experiments. Expressing the data in terms of $v_{\rm min}$ and $\tilde\eta(v_{\rm min})$, we can therefore compare the different experimental outcomes without any assumption about the dark halo of our galaxy. More precisely, we can separately compare the unmodulated and modulated parts  $\tilde\eta_0(v_{\rm min})=\sigma_p (\rho/m) \eta_0$ and $\tilde\eta_1(v_{\rm min})=\sigma_p (\rho/m) \eta_1$.

Most experiments do not measure the recoil energy $E$ directly, but rather a detected energy $E'$ (or a corresponding number of photoelectrons $N_{\rm pe}$) subject to measurement uncertainties and fluctuations. These are contained in an energy response function $G_{A,Z}(E,E')$ that incorporates the energy resolution $\sigma_E(E')$ and the mean value $\langle E'\rangle = E \, Q_{A,Z}(E)$, where $Q_{A,Z}(E)$ is the quenching factor (the analogous relation for photoelectrons is $\langle N_{\rm pe}\rangle = E \, {\mathcal Q}^{\rm pe}_{A,Z}(E)$). In this context, recoil energies are often quoted in keVnr, while detected energies are quoted in keVee (keV electron-equivalent) or  directly in photoelectrons. Moreover, experiments have an overall counting efficiency or cut acceptance $\epsilon(E')$ that depends on $E'$. A compound detector with mass fraction $C_{A,Z}$ in nuclide $A,Z$ has an expected event rate equal to
\begin{align}
\frac{dR}{dE'} = \epsilon(E') \, \int_0^\infty dE \, \sum_{A,Z} C_{A,Z} \, G_{A,Z}(E,E') \, \frac{dR_{A,Z}}{dE} .
\label{RSI-0}
\end{align}
Using $dE=(4\mu_{A,Z}^2/m_{A,Z}) v_{\rm min}  $ $dv_{\rm min}$, we can write the expected rate over a detected energy interval $[E'_1,E'_2]$ as
\begin{align}
R_{[E'_1,E'_2]} =  \int_0^\infty dv_{\rm min}  \,\, \mathcal{R}^{SI}_{[E'_1,E'_2]}(v_{\rm min})  \, \tilde\eta(v_{\rm min}) .
\label{RSI-2}
\end{align}
Here we have defined the response function for SI WIMP interactions, with \hfill\break $E_{A,Z} =2\mu_{A,Z}^2 v_{\rm min}^2 / m_{A,Z}$,
\begin{align}
\mathcal{R}^{SI}_{[E'_1,E'_2]}(v_{\rm min}) = \sum_{A,Z}  \frac{2\,v_{\rm min}\,C_{A,Z} \, \sigma^{SI}_{A,Z}(E_{A,Z})}{m_{A,Z}\,\sigma_p\,(E'_2-E'_1)}  \int_{E'_1}^{E'_2} dE' \,   G_{A,Z}(E_{A,Z} ,E')   \, \epsilon(E') .
\label{Response}
\end{align}
Notice that the temporal modulation of Eq.~\eqref{eta} produces a modulation of the rate
\begin{align}
R(t) = R_{0} + R_{1} \cos\!\left[ \omega (t-t_0) \right]
\end{align}
in any energy interval.

Our task is to gain knowledge on the functions $\eta_0(v_{\rm {min}})$ and $\eta_1(v_{\rm {min}})$ from measurements $\hat R_{0} \pm \Delta R_{0}$ and $\hat R_{1} \pm \Delta R_{1}$ of $R_{0}$ and $R_{1}$, respectively, in all the energy intervals probed by experiments. This is possible when a range of detected energies $[E'_1,E'_2]$ corresponds to only one range of $v_{\rm {min}}$ values  $[v_{\rm min,1},v_{\rm min,2}]$, for example when the measured rate is due to interactions with one nuclide only.   In this case, $[v_{\rm min,1},v_{\rm min,2}]$ is the $v_{\rm min}$ interval where  the response function $\mathcal{R}^{SI}_{[E'_1,E'_2]}(v_{\rm min})$ is significantly different from zero. We approximate this interval with $v_{\rm min,1} = v_{\rm min}(E'_1-\sigma_E(E'_1))$ and $v_{\rm min,2} = v_{\rm min}(E'_2+\sigma_E(E'_2))$ following Ref.~\cite{Frandsen:2011gi}.  When isotopes of the same element are present, like for Xe or Ge, the $v_{\rm min}$ intervals of the different isotopes almost completely overlap, and we take $v_{\rm min,1}$, $v_{\rm min,2}$ to be the $C_{A,Z}$-weighted averages over the isotopes of the element. When there are nuclides belonging to very different elements, like Ca and O in CRESST-II, a more complicated procedure should be followed (see Ref.~\cite{Gondolo:2012rs} for details). 

Once the $[E'_1,E'_2]$ range has been mapped to a $[v_{\rm min,1},v_{\rm min,2}]$ range, we can estimate the $v_{\rm min}$-weighted averages 
\begin{align}
\overline{\tilde\eta_{[E'_1,E'_2]}} = \frac{\int_{v_{\rm min,1}}^{v_{\rm min,2}} \mathcal{R}^{SI}_{[E'_1,E'_2]}(v_{\rm min}) \, \tilde\eta(v_{\rm min}) \, dv_{\rm min}}{\int_{v_{\rm min,1}}^{v_{\rm min,2}} \mathcal{R}^{SI}_{[E'_1,E'_2]}(v_{\rm min})  \, dv_{\rm min}} 
\end{align}
as
\begin{align}
  \overline{\tilde\eta_{[E'_1,E'_2]}}  = \frac{\hat{R}_{[E'_1,E'_2]}}{\mathcal{A}^{SI}_{[E'_1,E'_2]}},
\end{align}
where
\begin{align}
\mathcal{A}^{SI}_{[E'_1,E'_2]} = \int_{v_{\rm min,1}}^{v_{\rm min,2}} \mathcal{R}^{SI}_{[E'_1,E'_2]}(v_{\rm min})  \, dv_{\rm min} .
\end{align}

\section{Experiments considered}\label{sec:experiments}

In the following we compare direct detection data in a way similar to Ref.~\cite{Gondolo:2012rs} but  including new data: the new CDMS-II-Si results~\cite{Agnese:2013rvf},  a new measurement of the Na quenching factor~\cite{Collar:2013gu}, and a new measurement of the CoGeNT rate reduction  or acceptance factor~\cite{Aalseth:2012if}. In this section we summarize the data included in this analysis, stressing the elements which are new with respect to Ref.~\cite{Gondolo:2012rs}.

{\bf DAMA.} We read the modulation amplitudes from Fig.~6 of Ref.~\cite{Bernabei:2010mq}. We consider scattering off sodium only, since the iodine component is under threshold for low mass WIMPs and a reasonable local Galactic escape velocity. We show results for three values of the Na quenching factor: $Q_{\rm Na} = 0.30$, $Q_{\rm Na} = 0.45$ (suggested in Ref.~\cite{Hooper:2010uy}), and the energy dependent
$Q_{\rm Na,Collar}(E)$ obtained by  interpolating the central values of the data points  in  Fig.~9 of Ref.~\cite{Collar:2013gu}.  No channeling is included, as per Refs.~\cite{Bozorgnia:2010xy,Collar:2013gu}.

{\bf CoGeNT.} We use the list of events, quenching factor, efficiency, exposure times and cosmogenic background given in the 2011 CoGeNT data release \cite{CoGeNT2011release}. We bin data in 0.425 to 1.1125 keVee, 1.1125 to 1.8 keVee, 1.8 to 2.4875 keVee, and 2.4875 to 3.175  keVee energy intervals.
Differently from Ref.~\cite{Gondolo:2012rs}, we use here the acceptance shown in Fig.~20 of Ref.~\cite{Aalseth:2012if}, parametrized as $C(E) = 1 - \exp(- a E)$, with $E$ in keVee and $a = 1.32$ (``CoGeNT high"), $1.21$ (``CoGeNT med."), and $1.10$ (``CoGeNT low"). The uncertainty of this parametrization is much smaller than the uncertainty of the parametrization used in Ref.~\cite{Gondolo:2012rs}, which was modeled after Ref.~\cite{Kelso:2011gd}. As in \cite{Gondolo:2012rs}, in the figures we plot $\eta_0$ plus an unknown unmodulated background $b_0$ (which can vary from bin to bin).

{\bf CDMS-II.} As in Ref.~\cite{Gondolo:2012rs}, we use the germanium data (which we call CDMS-II-Ge) from the T1Z5 detector for the total rate~\cite{Ahmed:2010wy}, and the results presented in Ref.~\cite{Ahmed:2012vq}  for the modulation analysis. In addition, we also include the recent results from the silicon detector analysis in Ref.~\cite{Agnese:2013rvf}, which we call CDMS-II-Si. Since the energy resolution for silicon in CDMS-II has not been measured, we use the energy resolution for Ge 
in Eq.~(1) of Ref.~\cite{Ahmed:2009rh}, $\sigma(E) = \sqrt{0.293^2 + 0.056^2 E/{\rm keV}}$ keV. We corrected a mistake with respect to $\sigma(E)$ in Ref.~\cite{Gondolo:2012rs}, which however does not change the results of that paper. 
With three candidate events, we calculate the maximum gap upper limit~\cite{Yellin:2002xd} by taking $\eta(v_{\rm min})$ as a downward step function as in Ref.~\cite{Gondolo:2012rs}.
Assuming the events are a dark matter signal, we bin the recoil spectrum in 2 keVnr energy bins, resulting in 0 or 1 events per bin. In the energy bins containing one event we use the Poisson central confidence interval  of  (0.173, 3.30) expected events for zero background at the $68\%$ confidence level  to draw error bars. This is a conservative choice since the alternative Feldman-Cousins procedure to determine confidence levels would yields smaller error bars \cite{Feldman:1997qc}. We checked with the figure on page 16 of Ref.~\cite{Enectali} that the estimated background is less than $\sim 0.05$ events in each bin, thus it is negligible.

{\bf XENON100.} We use the last data release of Ref.~\cite{Aprile:2012nq}, with exposure of 224.6 days. As in Ref.~\cite{Gondolo:2012rs} for the energy resolution we use the Poisson fluctuation formula and a Gaussian single photoelectron resolution with $\sigma_{\rm PMT} = 0.5$ PE \cite{Aprile:2011hx}; however, our bound is now slightly different from the one derived in Ref.~\cite{Gondolo:2012rs} due to an error we found in the acceptance in our code. We use the acceptance shown in Fig.~1 of Ref.~\cite{Aprile:2012nq}, which is equivalent to setting $\mathcal{L}_{\rm eff}$ to 0 below $3$ keV. To be consistent with this choice we also cut $\mathcal{L}_{\rm eff}$ to 0 below $3$ keV for the first data release of Ref.~\cite{Aprile:2011hi}, with exposure of 48 kg $\times$ 100.9 days, in which case as shown in Fig.~\ref{Fig-Si} the limit obtained is always less stringent than the one obtained with the second data release of Ref.~\cite{Aprile:2012nq}. For this reason, we only adopt the bound from the latest data set.

{\bf XENON10.} We take the data from Ref.~\cite{Angle:2011th} and compute an upper limit following the procedure in Ref.~\cite{Gondolo:2012rs}. (We corrected a mistake in the computer program that affected the conversion from recoil energy to number of photoelectrons; after this correction, the limit is more stringent.) Again for the energy resolution we use the Poisson fluctuation formula and single photoelectron Gaussian resolution with $\sigma_{\rm PMT} = 0.5$ PE. The exposure is 1.2 kg $\times$12.5 days. We consider the $23$ events within the 1.4 keV-10 keV acceptance box in the  Phys.\ Rev.\ Lett.\ article (not the arXiv preprint).

{\bf SIMPLE.} We consider only the  Stage 2~\cite{Felizardo:2011uw}, a C$_2$ClF$_5$ detector with an exposure of 6.71 kg-day, one observed event above 8 keV, and an expected background of $2.2 \pm 0.30$ events. 
We use the Feldman-Cousins method \cite{Feldman:1997qc} to place an upper limit of 3.16 expected signal events for a 2.2 expected background and 1 observed event, at the $95\%$ confidence level.

{\bf CRESST-II.} We take the histogram of events in Fig.~11 of Ref.~\cite{Angloher:2011uu}. The electromagnetic background is modeled as one $e/\gamma$ event in the first energy bin of each module. The exposure is 730 kg-day. We assume a maximum WIMP velocity in the Galaxy such that W recoils can be neglected. A light WIMP will scatter however on both Ca and O, thus complicating the issue of associating to any range in detected energy $[E'_1, E_2']$ a range in $v_{\rm min}$ since the correspondence depends on the target mass. To do this, we follow the procedure described in Ref.~\cite{Gondolo:2012rs}.

\section{Results}

\begin{figure}[t]
\centering
\includegraphics[width=0.49\textwidth]{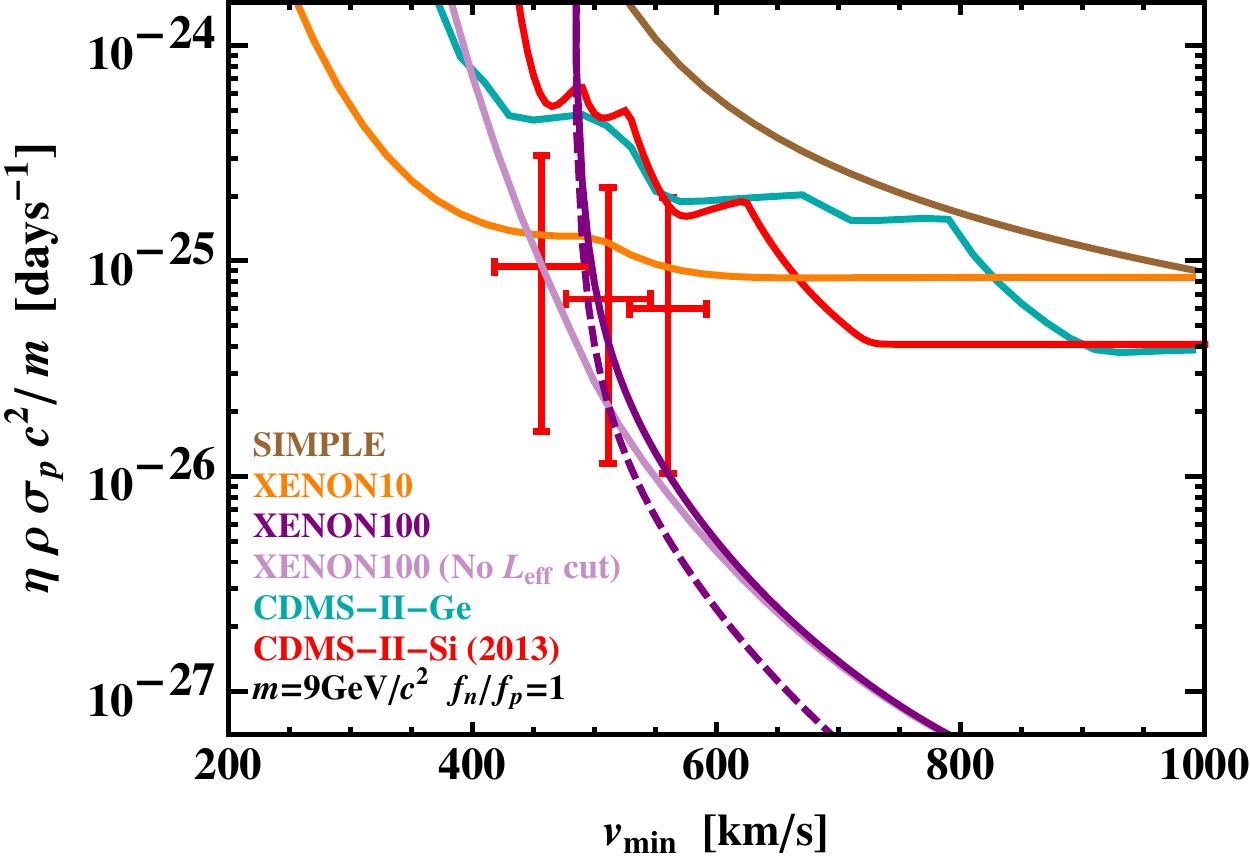}
\includegraphics[width=0.49\textwidth]{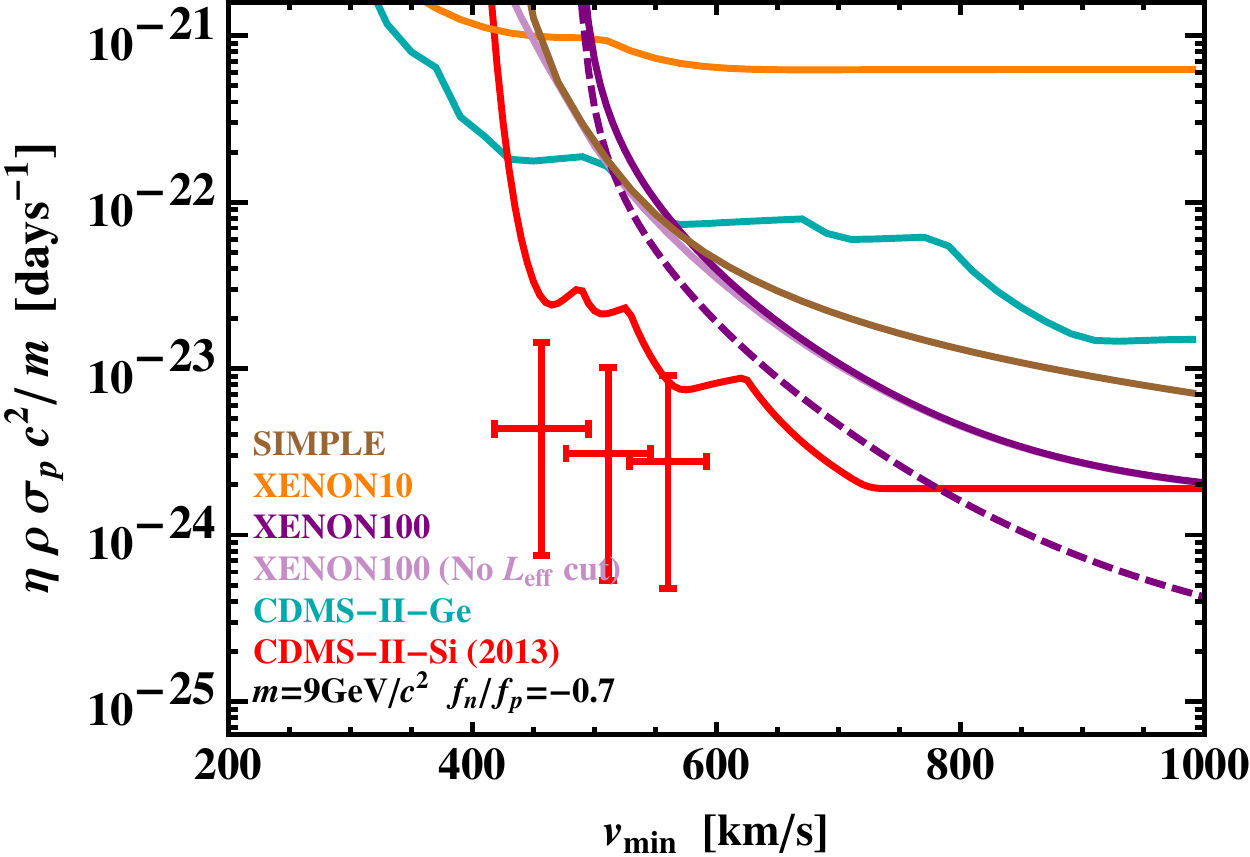}
\caption{CDMS-II-Si events (crosses), interpreted as a measurement of the unmodulated WIMP rate $\eta_0$, compared to the most stringent upper bounds on $\eta_0$,  for WIMPs with mass $m=$ 9 GeV/$c^2$, spin-independent interactions with nuclei, and (left) isospin-conserving $f_n/f_p = 1$ or (right) isospin-violating $f_n/f_p = -0.7$ couplings. 
For the isospin-conserving couplings, there is tension between the CDMS-II-Si events and the XENON100 limits, and a detailed statistical analysis would be necessary to quantify their degree of compatibility. For isospin-violating couplings, the  CDMS-II-Si events are compatible with all bounds (and the most stringent bound between 400 and 800 km/s in $v_{\rm min}$ is from the CDMS-II-Si data themselves). The three XENON100 lines indicate the 224.6 days bound (dashed) and the 100.9 days bound with and without $\mathcal{L}_{\rm eff}$ cut below 3 keV (dark and light solid line, respectively).}
\label{Fig-Si}
\end{figure} 

\begin{figure}[t]
\centering
\includegraphics[width=0.49\textwidth]{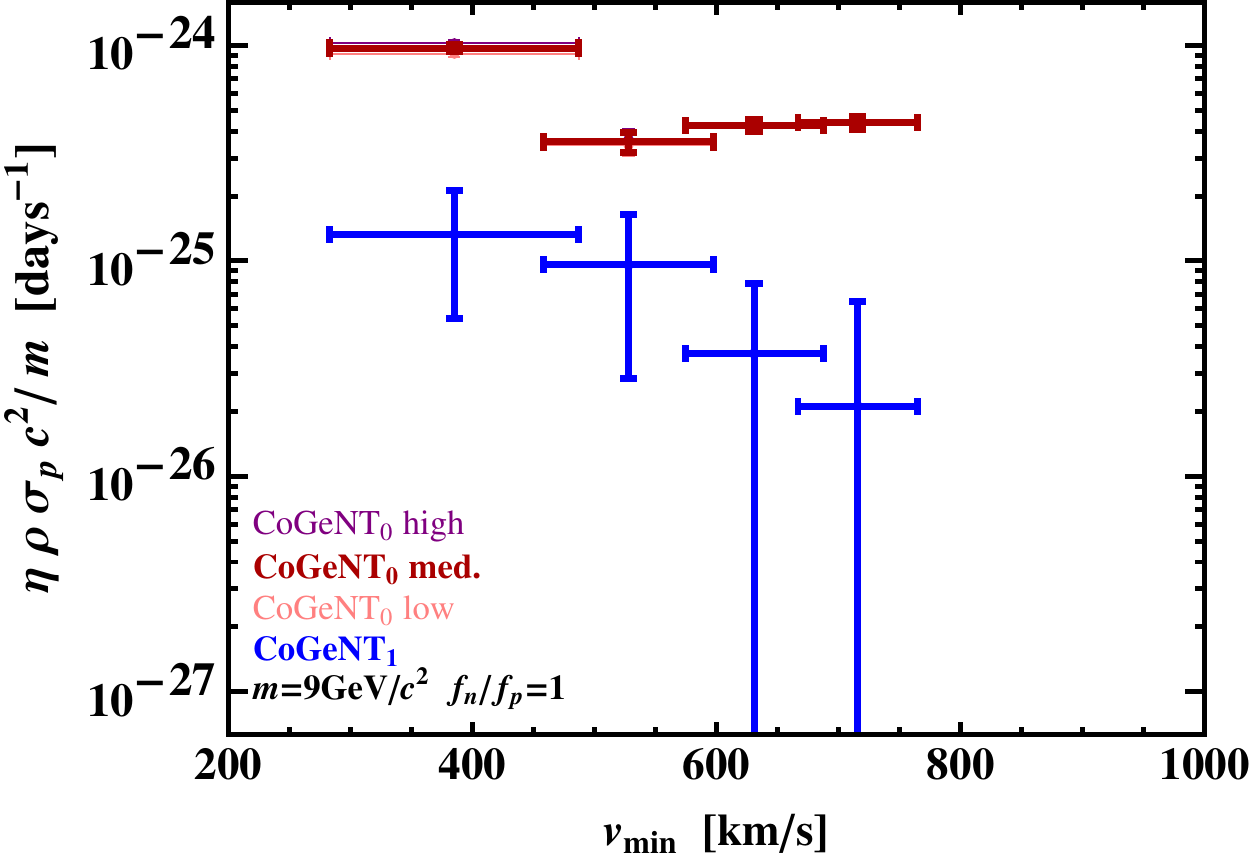}
\caption{CoGeNT measurements of the quantities $\eta_0+b_0$ and $\eta_1$ as functions of $v_{\rm min}$ for a WIMP with isospin-conserving couplings $f_n=f_p$ and mass $m=9$ GeV/$c^2$. Here $b_0$ is a constant background, $\eta_0$ is the unmodulated part of the velocity integral $\eta(v_{\rm min})$, and $\eta_1$ is its modulated part. The uncertainty in the acceptance, shown here with the three values of $\eta_0+b_0$ ``CoGeNT high", ``med.", and ``low", is visibly smaller than the uncertainty of the parametrization used in Ref.~\cite{Gondolo:2012rs}.}
\label{Fig-CoGeNT}
\end{figure}

\begin{figure}[t]
\centering
\includegraphics[width=0.49\textwidth]{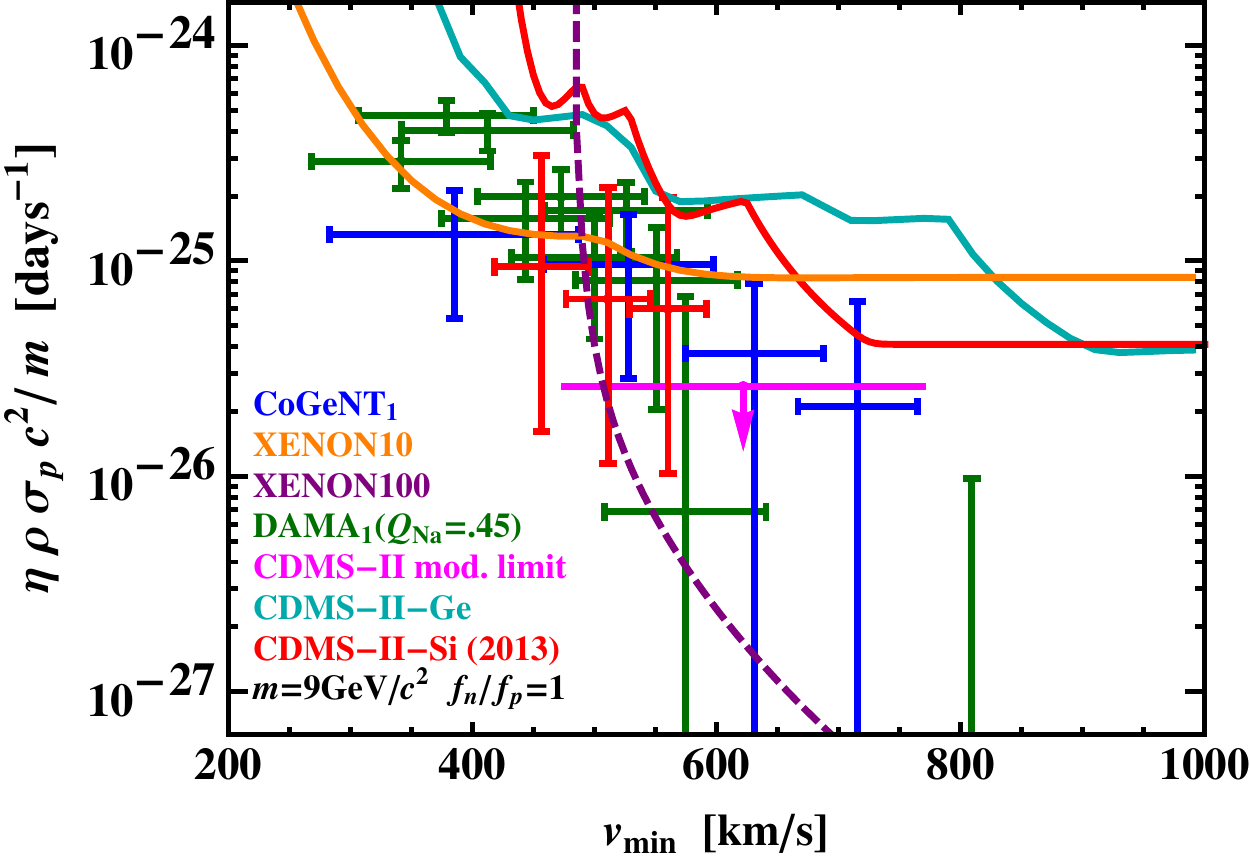}
\includegraphics[width=0.49\textwidth]{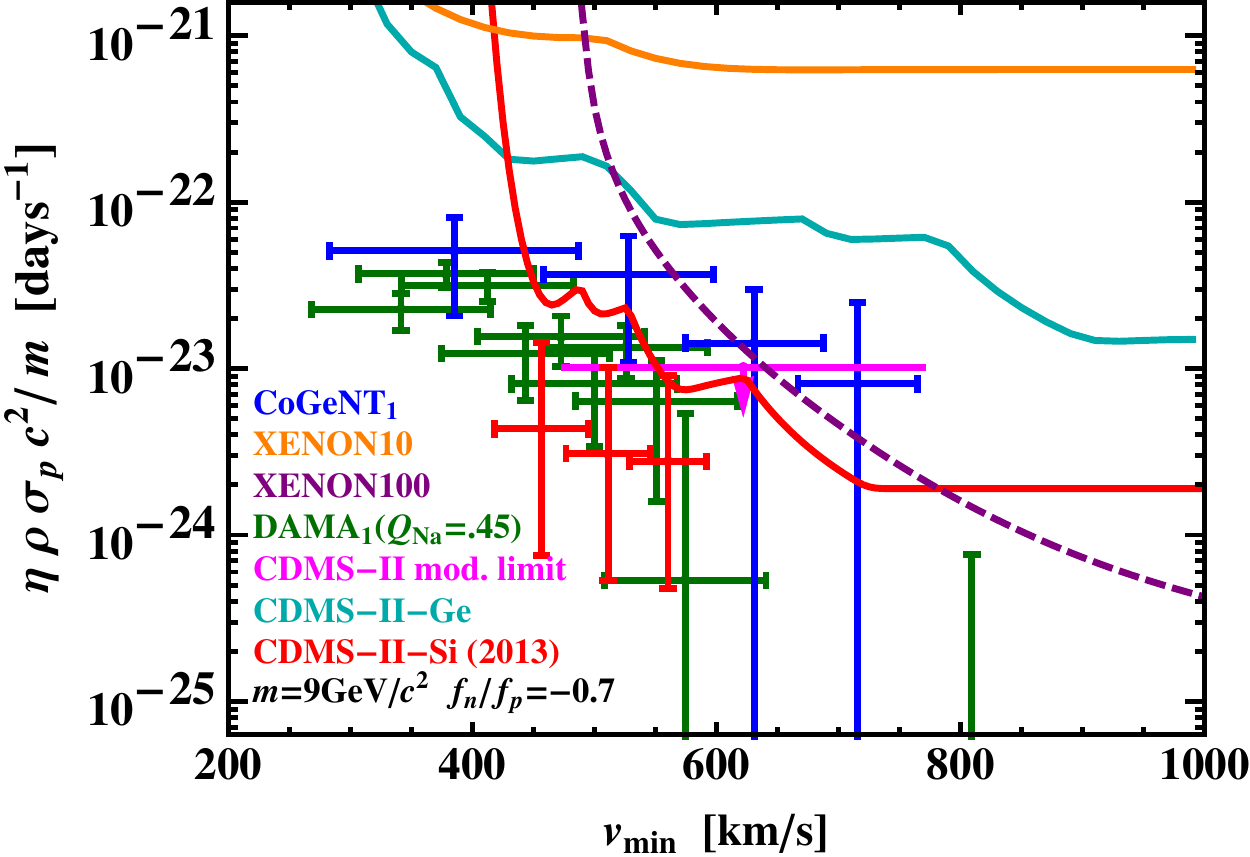}
\\
\includegraphics[width=0.49\textwidth]{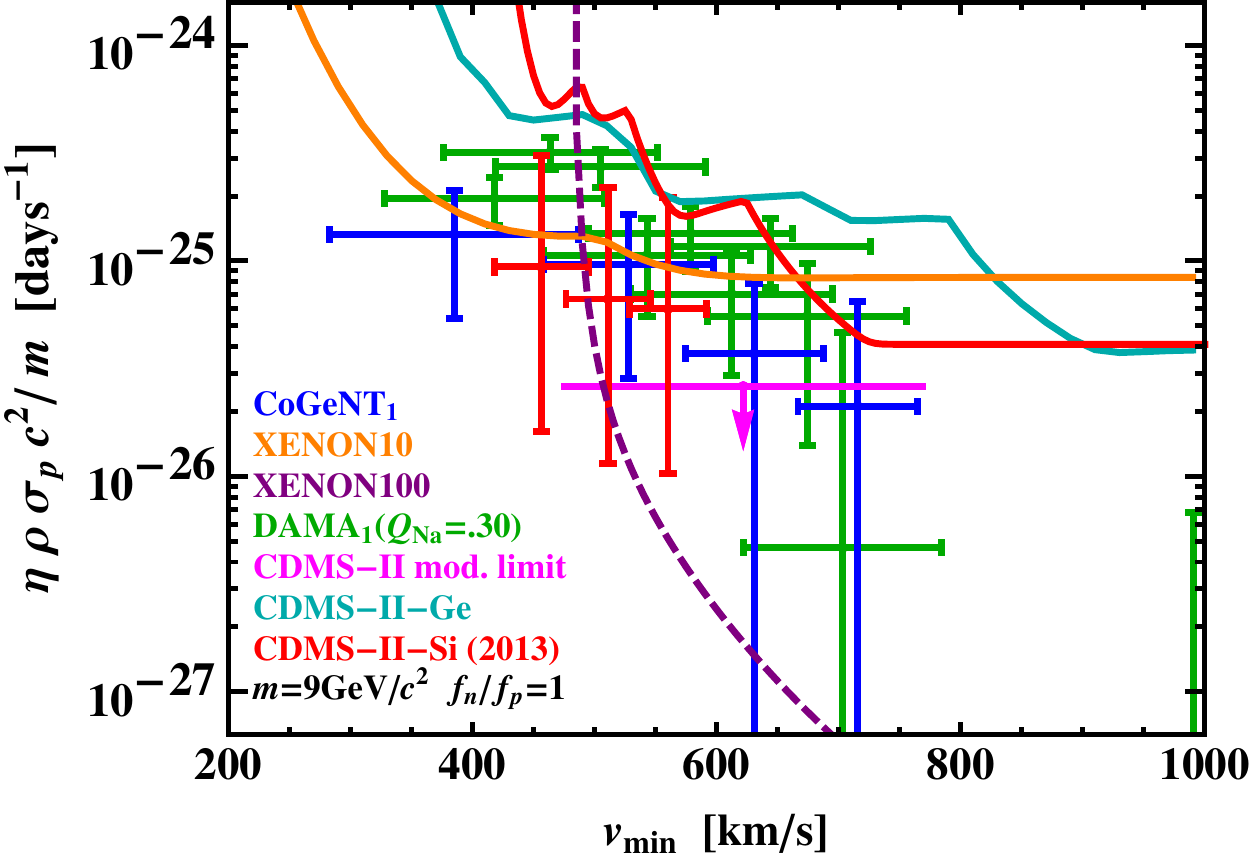}
\includegraphics[width=0.49\textwidth]{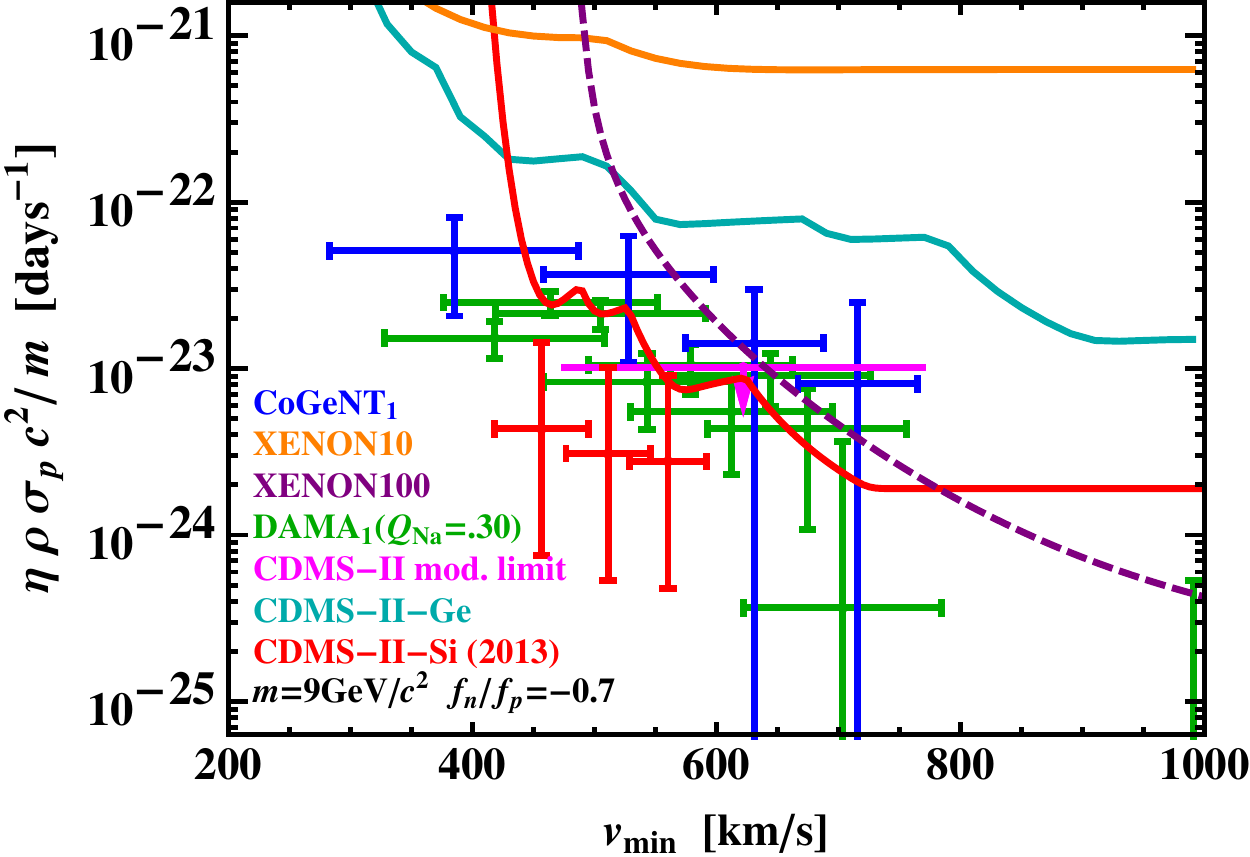}
\\
\includegraphics[width=0.49\textwidth]{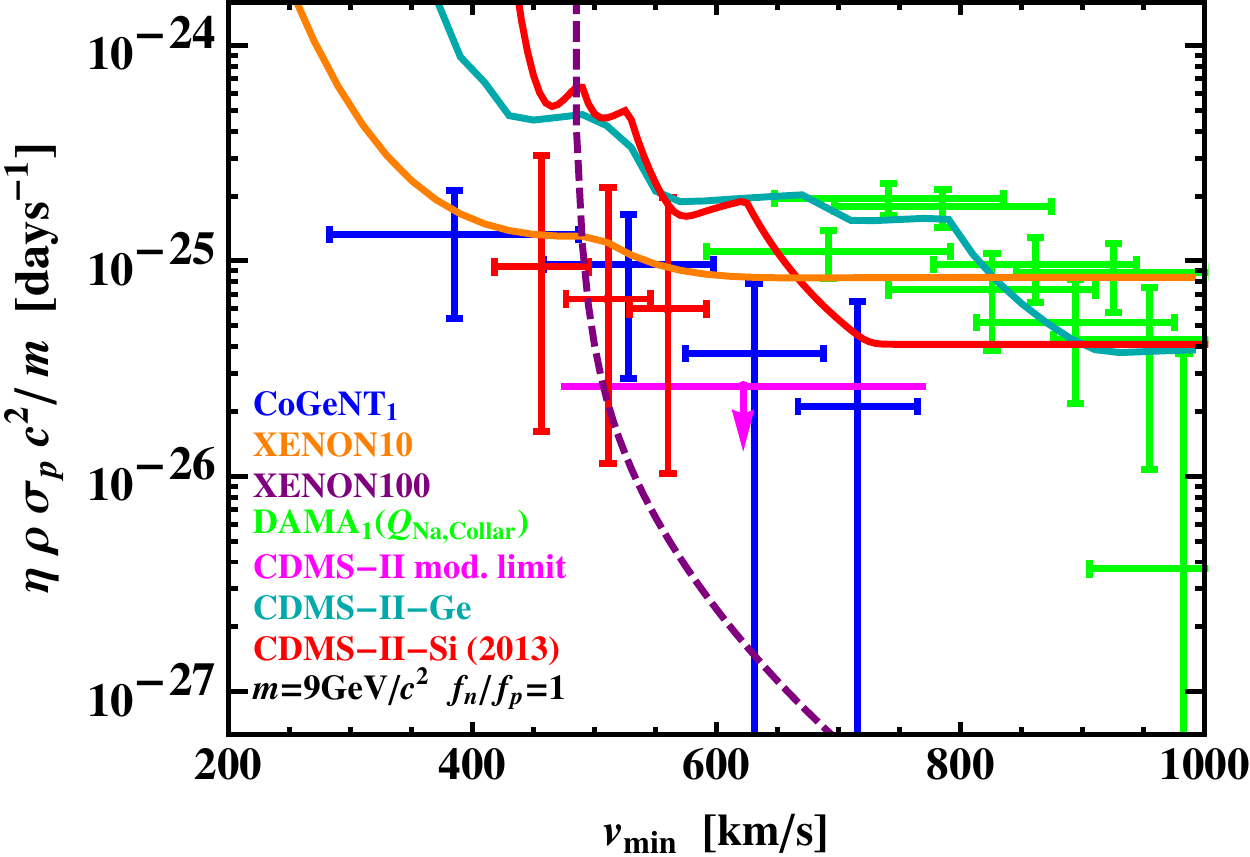}
\includegraphics[width=0.49\textwidth]{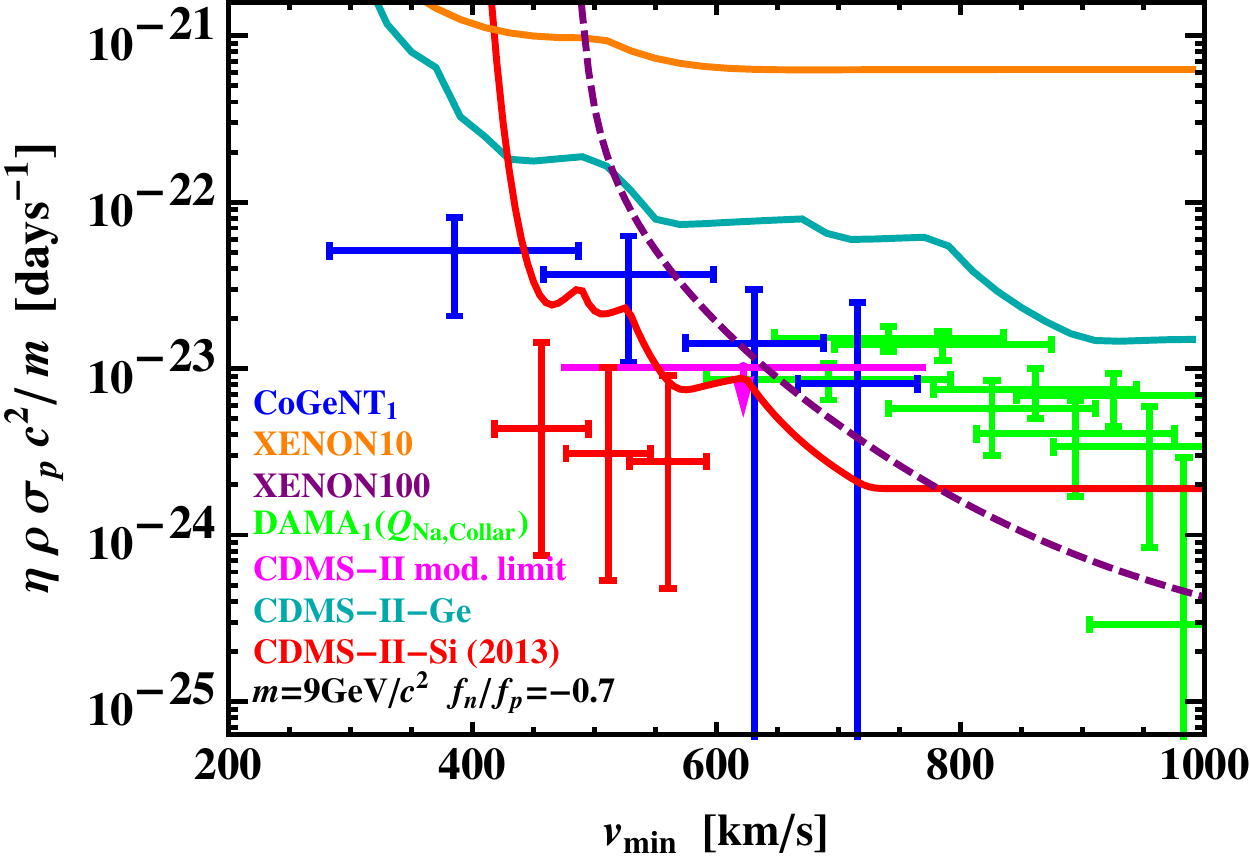}
\caption{Comparison of the modulated and unmodulated measurements of $\eta(v_{\rm min})$ as a function of  $v_{\rm min}$ for  $m=$ 9 GeV/$c^2$. The left column is for isospin-conserving couplings $f_n/f_p=1$; the right column is for isospin-violating couplings $f_n/f_p=-0.7$. The top row has sodium quenching factor in DAMA $Q_{\rm Na}=0.45$, the central row $Q_{\rm Na}=0.30$, and the bottom row $Q_{\rm Na}=Q_{\rm Na,Collar}(E)$ of Ref.~\protect\cite{Collar:2013gu}. The crosses and lines represent: for the modulated part $\eta_1$, the CoGeNT measurements (blue crosses), the DAMA measurements (green crosses), and the CDMS-II-Ge bound (magenta line with downward arrow); for the unmodulated part $\eta_0$, the CDMS-II-Si measurements (red crosses),  the CDMS-II-Ge bound (blue line), the CDMS-II-Si bound (red line), the XENON10 bound (orange line), and the XENON100 bounds (purple lines, dashed for the latest data). }
\label{Fig-three-Q}
\end{figure}

\begin{figure}[t]
 % \centering
\includegraphics[width=0.49\textwidth]{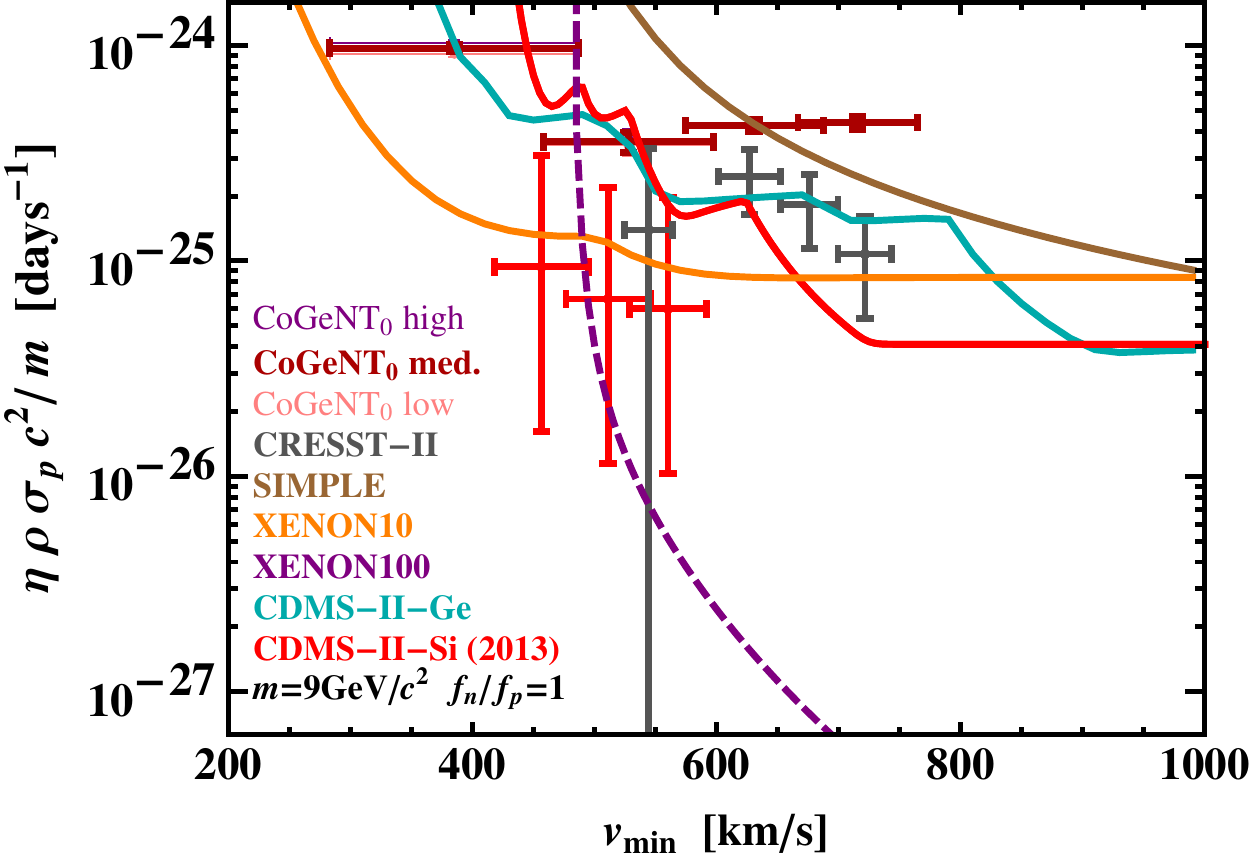}
\includegraphics[width=0.49\textwidth]{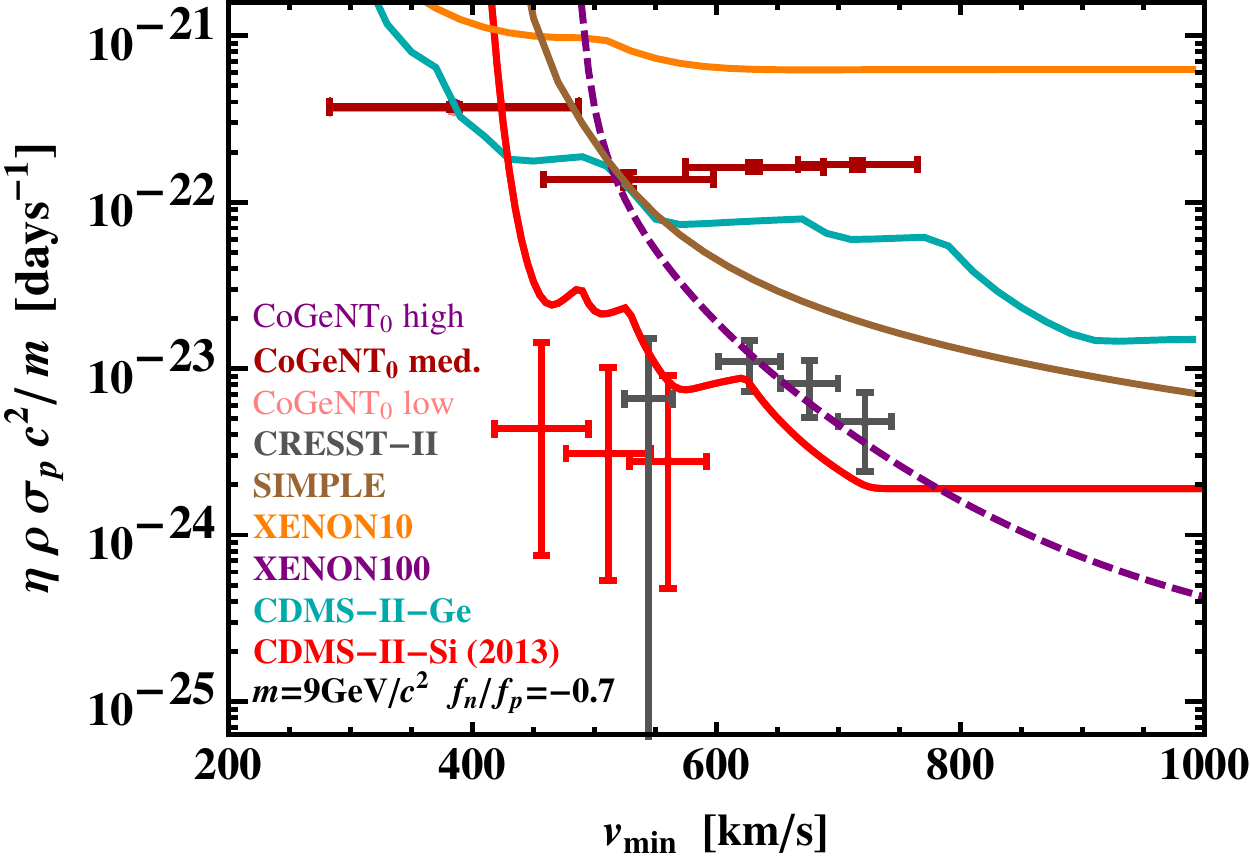}
\\
\includegraphics[width=0.49\textwidth]{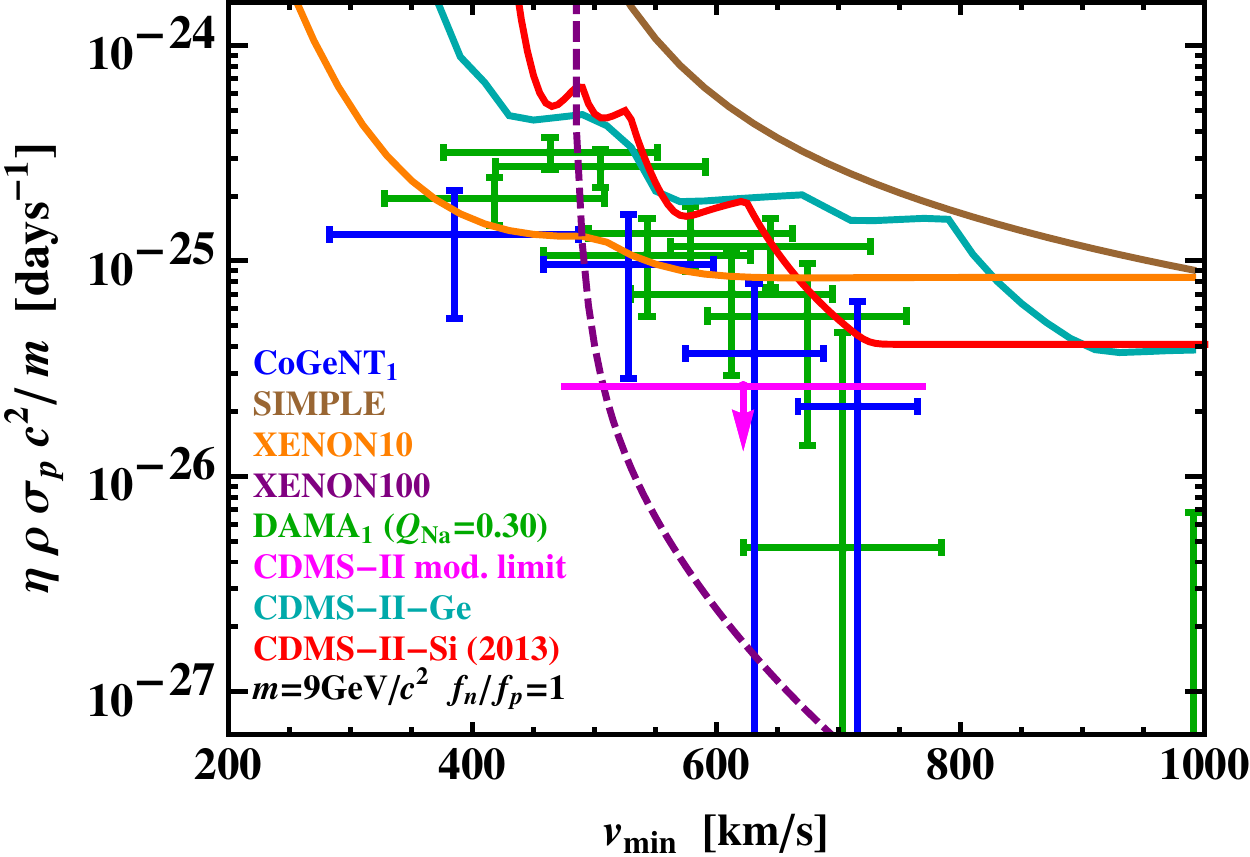}
\includegraphics[width=0.49\textwidth]{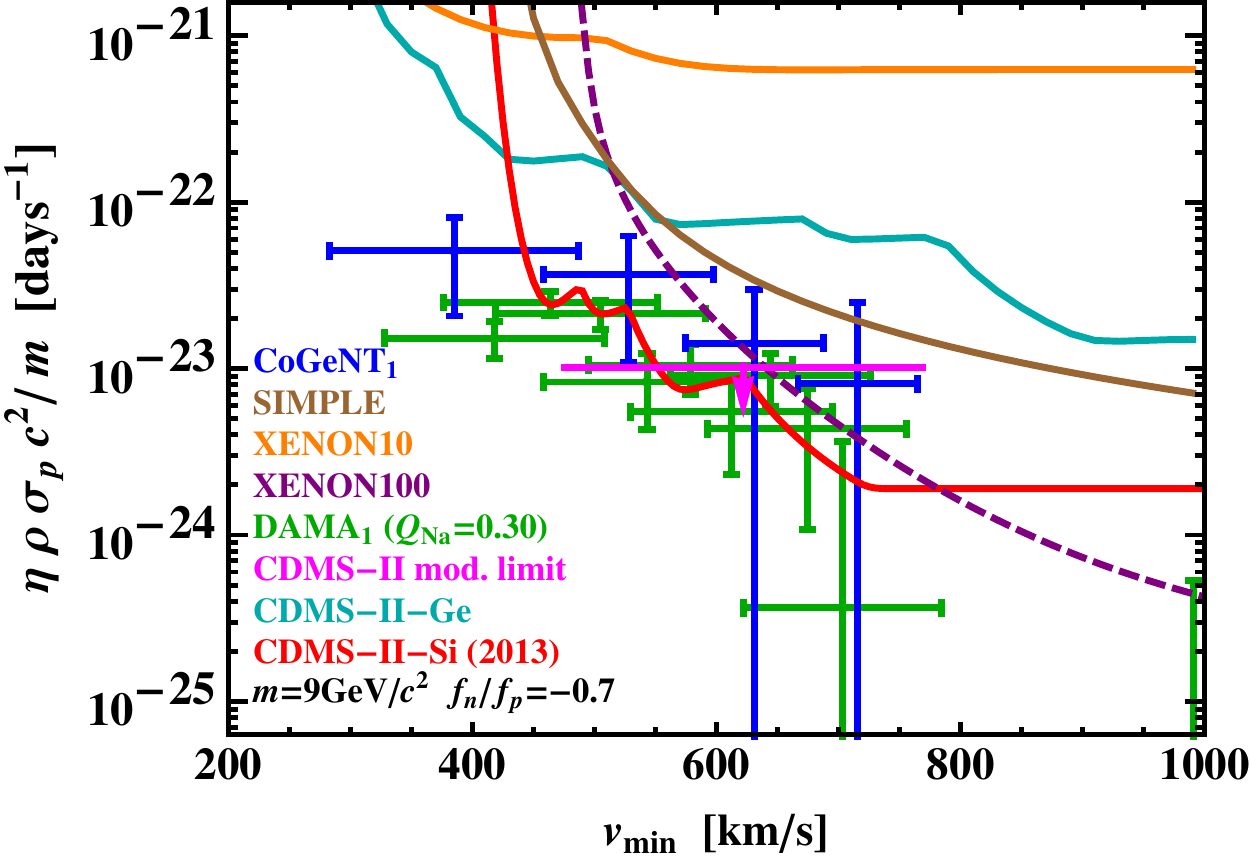}
\\
\includegraphics[width=0.49\textwidth]{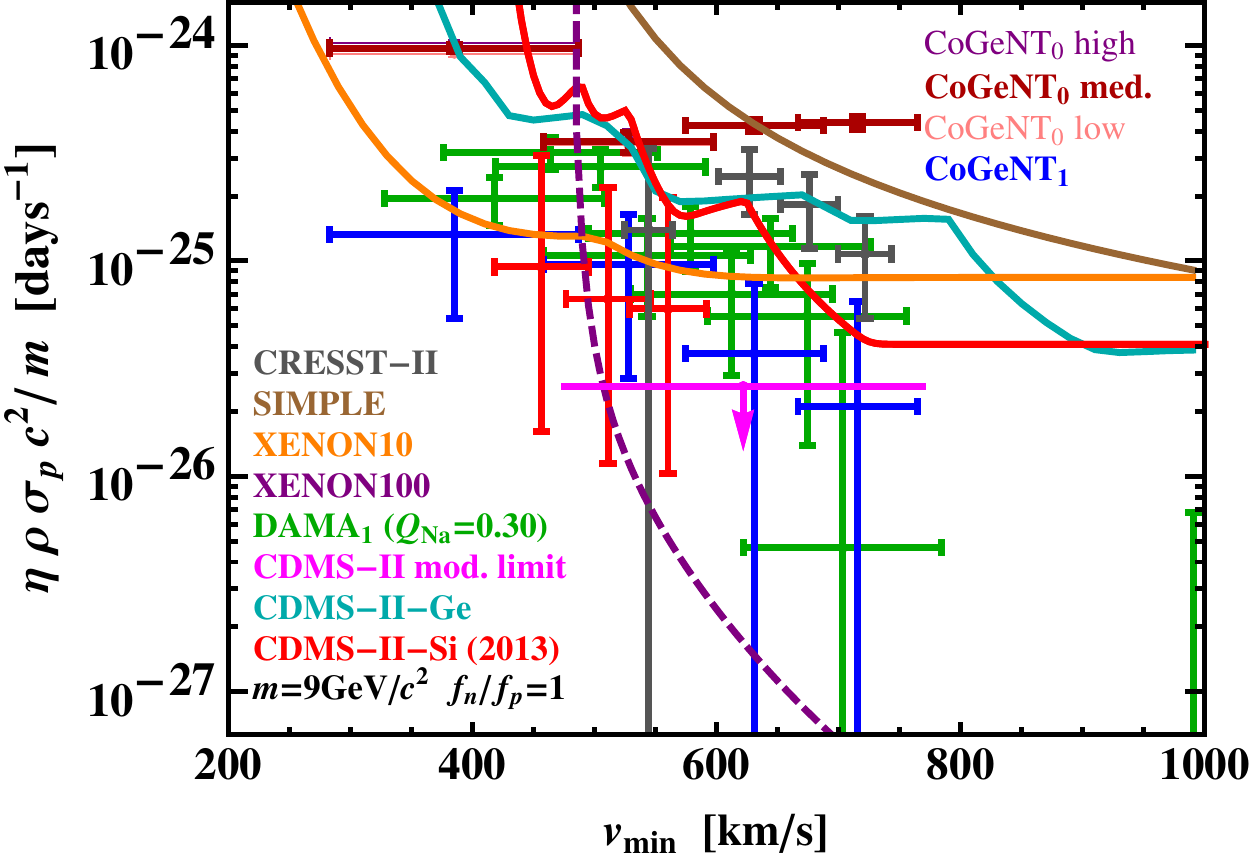}
\includegraphics[width=0.49\textwidth]{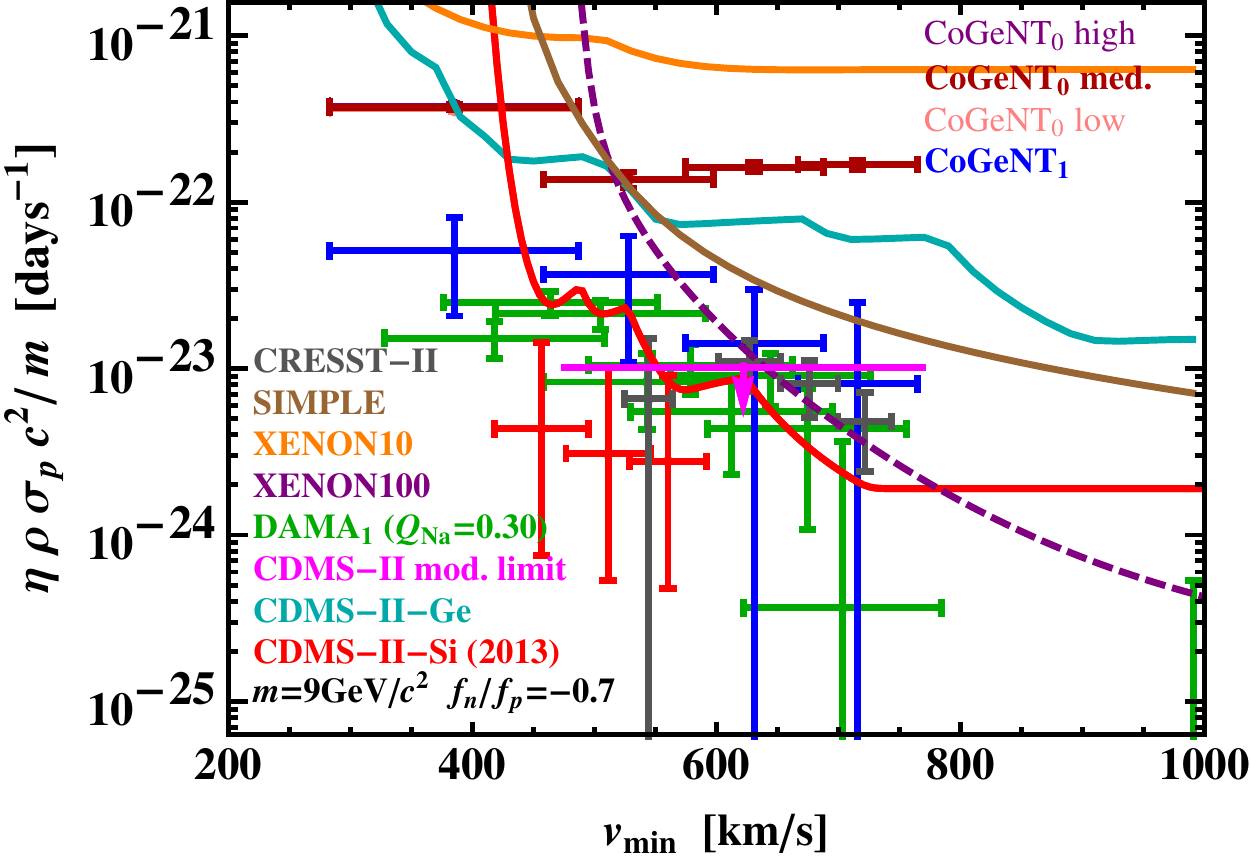}
\caption{
Top: bounds (XENON100, XENON10, CDMS-II-Ge, CDMS-II-Si, and SIMPLE) and measurements (CoGeNT, CRESST-II, and CDMS-II-Si) of the unmodulated DM signal $\tilde{\eta}_0(v_{\rm min})$. Middle: bounds (CDMS-II-Ge) and measurements (DAMA, CoGeNT) of the modulated component of the DM signal $\tilde{\eta}_1(v_{\rm min})$; the bounds on the unmodulated signal are also shown, as in the top panels. Bottom: All the measurements and bound on both $\tilde{\eta}_0(v_{\rm min})$ and $\tilde{\eta}_1(v_{\rm min})$, collected in a single figure. In these plots we assumed $Q_{\rm Na} =0.30$, $m=9$ GeV/$c^2$, and (left) $f_n/f_p = 1$, (right) $f_n/f_p = -0.7$. In the left panels, the XENON10 and XENON100 bounds exclude all but possibly the lowest energy CoGeNT and DAMA bin. For the candidates considered, the CDMS-II-Si measurements (red crosses) either overlap (left) or are lower than (right) the CoGeNT and DAMA measurements of the modulated part (blue and green crosses, respectively). This makes it difficult to interpret the CDMS-II-Si events as the unmodulated part of a WIMP signal, since the modulated part is expected to be smaller than the unmodulated part.}
\label{Fig-ALL-9GeV}
\end{figure}

\begin{figure}[t]
%  \centering
\includegraphics[width=0.49\textwidth]{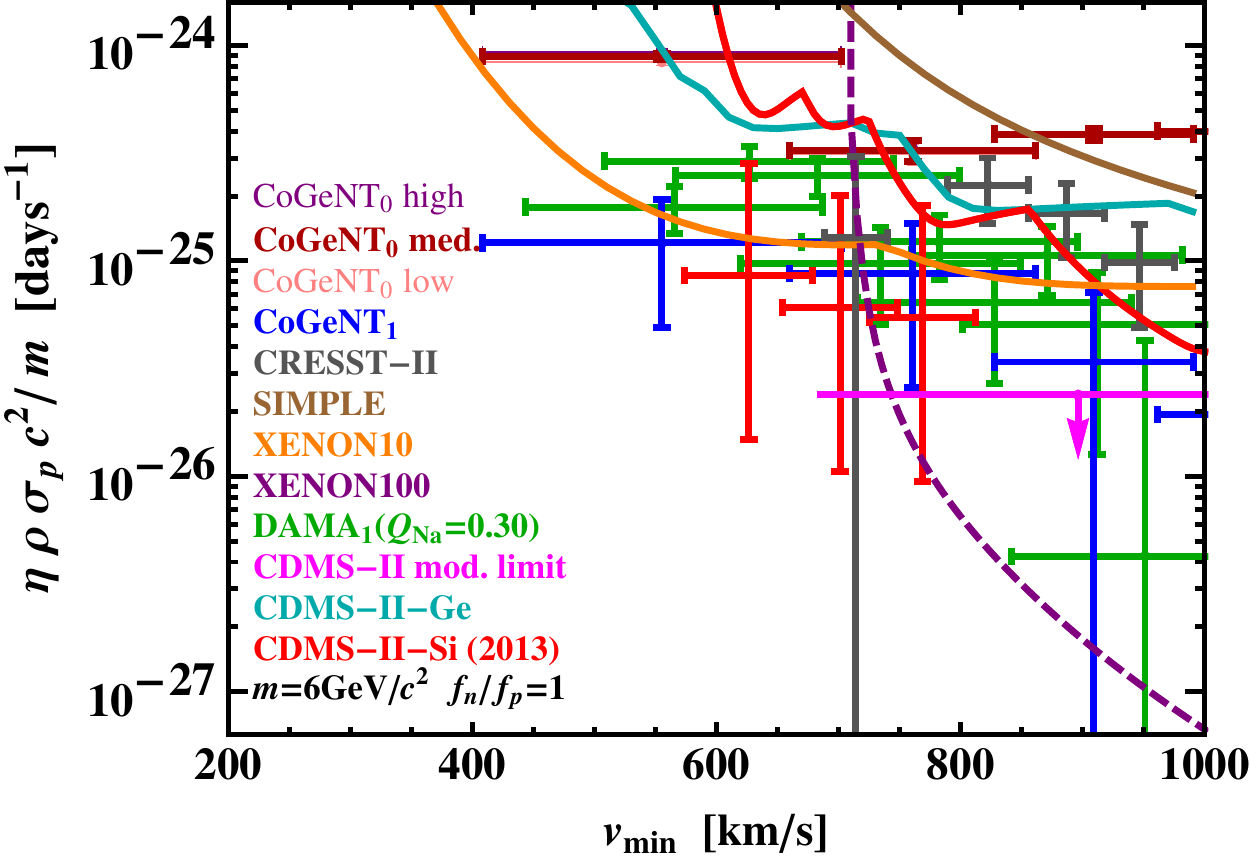}
\includegraphics[width=0.49\textwidth]{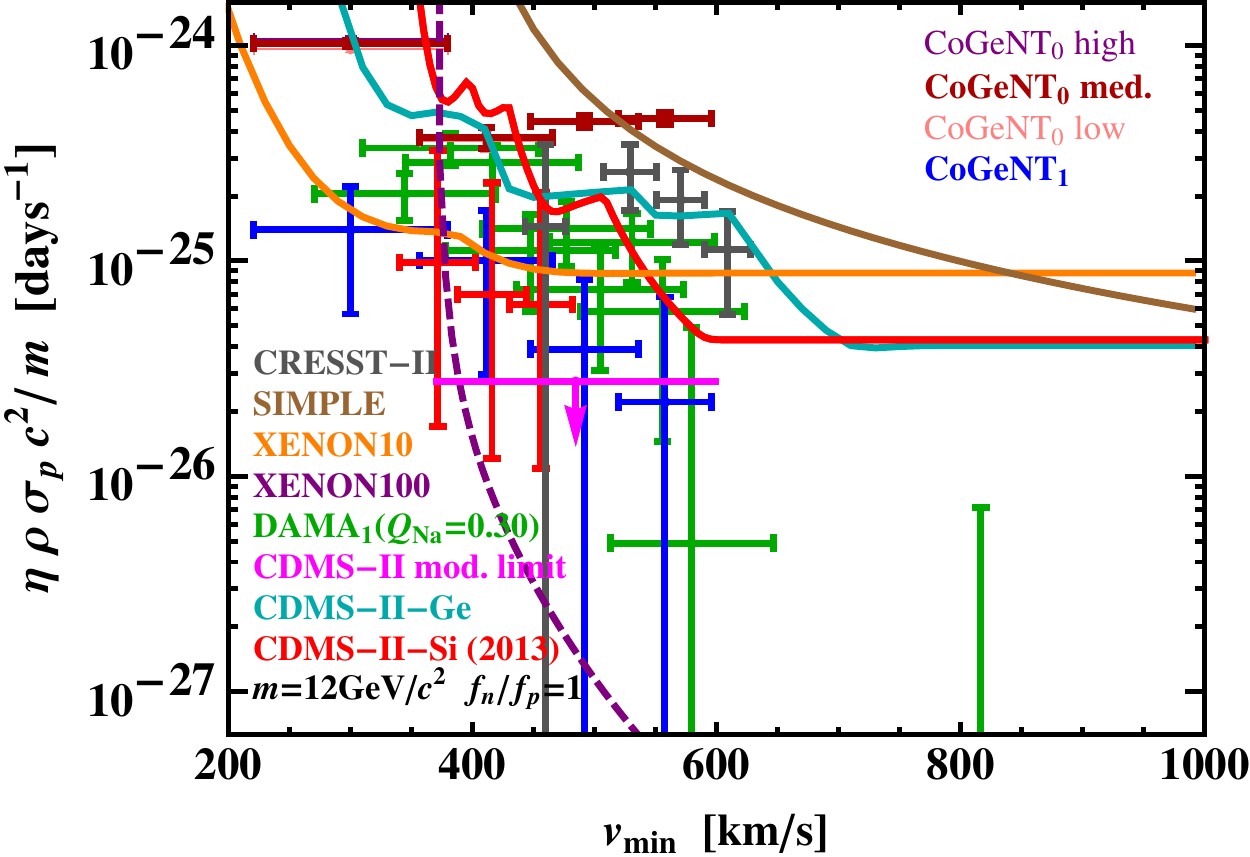}
\caption{As in Fig.~\protect\ref{Fig-ALL-9GeV}, but for  (left) $m=$ 6 GeV/$c^2$ and  (right) $m=$ 12 GeV/$c^2$. The main results are qualitatively the same as for $m=$ 9 GeV/$c^2$ in Fig.~\protect\ref{Fig-ALL-9GeV}.}
\label{Fig-ALL-6and12GeV}
\end{figure}

Figs.~\ref{Fig-Si} to~\ref{Fig-ALL-6and12GeV} show our results for a WIMP with spin-independent interactions and either isospin-conserving couplings ($f_n/f_p=1$) or isospin-violating couplings ($f_n/f_p=-0.7$).  In all the figures, the vertical axis shows the quantity $\tilde{\eta}c^2$, which has units of inverse days. Despite these units, $\tilde\eta c^2$  is not the number of WIMPs impinging on the detector per day. Notice that $\eta$ in the label of the vertical axis stands for either $\eta_0$ or $\eta_1$ depending on the experiment. In many cases we plot measurements or bounds for both the unmodulated, $\eta_0$, and modulated, $\eta_1$, parts of $\eta$ in the same figure to be able to compare them. In all realistic cases one should have $\eta_1$ sufficiently smaller than $\eta_0$.

In Fig.~\ref{Fig-Si} we consider the CDMS-II-Si data points as a measurement of  the unmodulated rate $\eta_0$, and we compare them to the most stringent upper bounds on $\eta_0$\,  for WIMPs with mass  $m=$ 9 GeV/$c^2$ and spin-independent interactions. For XENON100 we show both the 224.6 days bound of Ref.~\cite{Aprile:2012nq} (dashed line), and the 100.9 days bound of Ref.~\cite{Aprile:2011hi}, with and without $\mathcal{L}_{\rm eff}$ cut below 3 keV (dark and light purple line, respectively). The left panel has $f_n/f_p=1$, the right panel has $f_n/f_p=-0.7$. For the isospin-conserving couplings, there is tension between the CDMS-II-Si events and the XENON10 and XENON100 limits, and a detailed statistical analysis would be necessary to quantify their degree of compatibility. For isospin-violating couplings, the CDMS-II-Si events are compatible with all bounds (and the most stringent bound between 400 and 800 km/s in $v_{\rm min}$ is from the CDMS-II-Si data themselves).

Fig.~\ref{Fig-CoGeNT} shows the CoGeNT measurement of (i) the unmodulated part $\eta_0$ of the velocity integral $\eta(v_{\rm min})$ plus a constant background $b_0$, i.e.~$\eta_0 +b_0$, and (ii) the modulated part $\eta_1$ of $\eta(v_{\rm min})$. Both are given as functions of  $v_{\rm min}$ for a WIMP with isospin-conserving couplings and mass $m=$ 9 GeV/$c^2$. Three lines are drawn for each $\eta_0+b_0$ cross: one for the central value of the acceptance in Fig.~20 of Ref.~\cite{Aalseth:2012if}  (``CoGeNT med."), one for the high value of the acceptance (``CoGeNT high"),  and one for the low value of the acceptance (``CoGeNT low"). In comparison with the analogous figure in Ref.~\cite{Gondolo:2012rs}, Fig.~\ref{Fig-CoGeNT} still shows that the modulation amplitude of the CoGeNT data (in blue) is large with respect to the average (in brown), certainly larger than the few percent modulation in usual halo models.  The differences between the high, low and medium $\eta_0$ measurements are now however hardly distinguishable, contrary to the situation in Ref.~\cite{Gondolo:2012rs}, because the uncertainty in the acceptance factor of CoGeNT in Ref.~\cite{Aalseth:2012if}  is now smaller than what it was in Ref.~\cite{Gondolo:2012rs}.

In Ref.~\cite{Frandsen:2011gi}, a bound on the modulation fraction has been inferred within the conventional halo model and within models inspired by numerical simulations of dark halo formation without baryons (in particular those in Ref.~\cite{Kuhlen:2009vh}). In these models, the ratio $\eta_1 / \eta_0$ is less than $\sim 10\%$ at   $v_{\rm {min}} \sim 200- 300$ km/s, and less than $\sim 40\%$ at  $v_{\rm {min}} \sim 300- 500$ km/s, while the CoGeNT data in these $v_{\rm {min}}$ ranges show typical values of about 10\% and 30\% for $m= 9$ GeV (see e.g.~Fig.~\ref{Fig-CoGeNT}). Since we do not separate the contribution of the background from the average rate, these values could also be larger. If they are larger, it  may mean that the CoGeNT modulation is not due to dark matter or that the halo model is different from those examined. Additional structures such as streams or debris flows can further enhance the modulation fraction, although a modulation fraction close to 100\% is usually expected only at very large values of  $v_{\rm {min}}$  (in Ref.~\cite{Frandsen:2011gi}  $\eta_1 / \eta_0$ reaches values of  $\sim 80\%$ at   $v_{\rm {min}} \sim 600$ km/s). An assessment of possible modulation fractions in non-conventional models requires a dedicated study beyond the scope of this paper. Thus we prefer to be conservative and make no assumptions on the modulation fraction, in the same way as we do not make any assumption on the dark matter halo.

In Fig.~\ref{Fig-three-Q} we compare the measurements of the modulated part of $\eta(v_{\rm min})$ for  $m=$ 9 GeV/$c^2$ with the CDMS-II-Si data points and the most stringent modulated and unmodulated upper bounds. For the modulated part $\eta_1$ we show the CoGeNT measurements (blue crosses), the DAMA measurements (green crosses), and the CDMS-II-Ge bound (magenta line with downward arrow). For the unmodulated part $\eta_0$ we show the CDMS-II-Si measurements (red crosses),  the CDMS-II-Ge bound (blue line), the CDMS-II-Si bound (red line), the XENON10 bound (orange line), and the XENON100 bounds (purple lines, dashed for the latest data). The three plots on the left column have isospin-conserving couplings $f_n/f_p=1$. The three plots on the right column have isospin-violating couplings $f_n/f_p=-0.7$. The rows differ in the choice of the sodium quenching factor $Q_{\rm Na}$ for the DAMA experiment: the first row has the large $Q_{\rm Na}=0.45$ of Ref.~\cite{Hooper:2010uy}, the second row has the  medium $Q_{\rm Na}=0.30$ used by the DAMA/LIBRA collaboration, and the third row has the low and energy-dependent  $Q_{\rm Na,Collar}(E)$ of Ref.~\cite{Collar:2013gu}. As $Q_{\rm Na}$ decreases, the DAMA points (green) move to higher $v_{\rm min}$ (to the right). With the lowest values of $Q_{\rm Na,Collar}(E)$, close to 0.10 at the lowest energies, the higher energy data points of DAMA would be incompatible with reasonable values of the Galactic escape velocity (for light WIMPs), and all the points are rejected by several upper bounds.

The case of spin-independent isospin-conserving couplings (the left column of Fig.~\ref{Fig-three-Q}) is similar to that in Ref.~\cite{Gondolo:2012rs}: only the lowest energy bins of the CoGeNT modulation data and of the DAMA modulation data with $Q_{\rm Na} = 0.45$ may escape the XENON10 and XENON100 bounds and the CDMS-II-Ge modulation bound. The situation for the recent CDMS-II-Si events appears similar, although the error bars are larger. The CDMS-II-Si data points are in the region partially excluded by the XENON100 and XENON10 bounds, although the error bars of the two lowest energy bins extend into the allowed region. In this spin-independent isospin-conserving case, the unmodulated part $\eta_0$ measured by CDMS-II-Si overlaps with the modulated part $\eta_1$ measured by CoGeNT and/or DAMA. This makes it difficult to interpret the CoGeNT and DAMA annual modulations on one hand and the CDMS-II-Si events  on the other as the modulated and unmodulated aspects of the same WIMP dark matter signal.

For the case of  isospin-violating couplings with $f_n/f_p \simeq -0.7$ (right column of Fig.~\ref{Fig-three-Q}),  the CDMS-II-Si upper bounds are now the most restrictive, having taken the role of the XENON100 bounds. This was to be expected, since the particular ratio $f_n/f_p=-0.7$ was chosen to decrease the WIMP-Xe couplings. Notice that also the WIMP coupling with the other elements decrease, and thus $\tilde\eta$ increases by the same amount. The rescaling factor between the isospin-violating and the isospin-conserving case is $A^2 / [Z + (A-Z)(f_n/f_p)]^2$, where with $f_n/f_p=-0.7$. This factor is $\sim400$ for CDMS-II-Ge and CoGeNT, $\sim50$ for CDMS-II-Si, $\sim80$ for DAMA-Na, $\sim8000$ for XENON, and $\sim50$ for CRESST-II (considering only scattering off Ca and O). Therefore, for example, for $f_n/f_p=-0.7$ the DAMA modulation points are below the CoGeNT data points, while they are above  for $f_n/f_p=1$. With isospin-violating couplings the most severe bounds on the DAMA and CoGeNT modulation amplitudes come from the CDMS-II modulation limit and the CDMS-II-Si bound. With respect to the previous analysis in Ref.~\cite{Gondolo:2012rs}, there is now a slight tension between the modulated and unmodulated data also in the case of isospin-violating couplings (right column in Fig.~\ref{Fig-three-Q}). The new CDMS-II-Si bounds (red line) is lower than the central values of all but the lowest energy modulated CoGeNT data (blue crosses), although the CoGeNT error bars extend well into the allowed region. While pushing the XENON10 and XENON100 bounds away from the modulated data, the $f_n/f_p=-0.7$ isospin-violation pushes at the same time the CDMS-II-Si events below the CoGeNT and DAMA modulated data. This makes it difficult, as it was for the isospin-conserving scenario, to interpret the CDMS-II-Si events and the modulated direct detection data as coming from spin-independent interactions of dark matter WIMPs (unless the CDMS-II-Si events are themselves part of a modulated signal).


Fig.~\ref{Fig-ALL-9GeV} shows all the measurements and bounds on the modulated and unmodulated parts of $\eta(v_{\rm min})$ as a function of  $v_{\rm min}$ included in this analysis: the  CoGeNT and  DAMA  measurements  of, and the CDMS-II-Ge bounds on, the modulated part $\eta_1$;  the CDMS-II-Si, CoGeNT and CRESST-II measurements, and the CDMS-II, XENON10, XENON100 and SIMPLE bounds on the unmodulated part $\eta_0$. In both left and right panels $Q_{\rm Na} = 0.30$ and $m=$ 9 GeV/$c^2$.  It is assumed that the WIMP couplings are isospin-conserving in the left panel an isospin-violating with $f_n/f_p = -0.7$ in the right panel. For isospin-conserving coupling the XENON10 bound seems to completely exclude the $\eta_0 + b_0$ measurement of CoGeNT. The tension between the two may be alleviated with a smaller $\eta_0$ provided it would not conflict with the CoGeNT $\eta_1$ measurement. The CRESST-II measurement of the unmodulated rate is incompatible with the CoGeNT and DAMA  measurements of the modulated rate for the WIMP candidates considered, since the CRESST-II  data points either  fall above (left) or overlap (right) the CoGeNT and DAMA modulation data.

The last figure, Fig.~\ref{Fig-ALL-6and12GeV},  shows again all the measurements and bounds on the modulated and unmodulated parts of $\eta(v_{\rm min})$ included in this analysis, but for different WIMP masses than in the previous figures, $m= 6$ GeV/$c^2$ in the left panel and  $m= 12$ GeV/$c^2$ in the right panel.  In both panels we assume isospin-conserving WIMP-nucleon couplings. The main results are qualitatively the same as for $m=$ 9 GeV/$c^2$. However, we notice that more of the CDMS-II-Si crosses come out of the region rejected by the XENON100 bound as the WIMP mass decreases (while the XENON10 bound remains almost equally stringent). At $m= 6$ GeV, not all of the CDMS-II-Si crosses escape the XENON100 limit (see Fig.~\ref{Fig-ALL-6and12GeV}a). This plot does not make a clear case for compatibility, in spite of the fact that analyses in the $m$--$\sigma$ plane \cite{Agnese:2013rvf,Frandsen:2013cna} suggest a $\sim90$\% compatibility around $m\sim 7$ GeV.
When $m$ is much smaller than the masses of all  nuclei involved,  the combination $m v_{\rm min}$ becomes independent of $m$. Thus in this limit all data points and bounds presented in the figures scale together in $v_{\rm min}$ by  the ratio $(9 ~{\rm GeV}/mc^2)$. For the target nuclei considered, this argument applies rather well  to WIMP masses smaller than 9 GeV/$c^2$, with some small deviations for the lightest nuclei (O, Na and Si), as a careful examination of Fig.~\ref{Fig-ALL-6and12GeV} (left) shows. The scaling argument is still  approximately correct even for somewhat larger masses, such as $m=$ 12 GeV/$c^2$.  All the data points and limits in  Fig.~\ref{Fig-ALL-6and12GeV}  are approximately scaled towards lower or higher  $v_{\rm min}$  values with respect to their locations in Fig.~\ref{Fig-ALL-9GeV}, those of CRESST-II (oxygen) somewhat less than the others.

\section{Conclusions}

We have used the halo-independent method to compare the current direct detection data on light dark matter WIMPs ($\sim 1$-10 GeV/$c^2$ in mass). We have followed the specific method presented by two of us in Ref.~\cite{Gondolo:2012rs}, which takes into account form factors, experimental  energy resolutions, acceptances, and efficiencies with arbitrary energy dependence. We have focused on light WIMPs with spin-independent WIMP-nucleus interactions and with isospin-conserving and isospin-violating couplings to protons and neutrons, in the ratios $f_n/f_p=1$ and $f_n/f_p=-0.7$, respectively. 

Compared to Ref.~\cite{Gondolo:2012rs}, we have added the three events off silicon recently reported by the CDMS-II collaboration against an expected background of $0.7$ events~\cite{Agnese:2013rvf}, together with the upper limits implied by the same data. We have also updated the CoGeNT acceptance to its latest published form~\cite{Aalseth:2012if}, and have included the case of a very low sodium quenching factor $Q_{\rm Na}$ for the DAMA experiment as measured in Ref.~\cite{Collar:2013gu}. As expected, the low $Q_{\rm Na}$ makes the DAMA modulation data incompatible with a reasonable escape velocity for light WIMPs, and all the DAMA points are rejected by several upper bounds.

For isospin-conserving couplings, similarly to Ref.~\cite{Gondolo:2012rs}, we conclude that only the lowest energy bins of the CoGeNT and DAMA modulation data (the latter with $Q_{\rm Na} = 0.45$) may escape the XENON10, XENON100 and the CDMS-II-Ge modulation bounds. For isospin-violating couplings  $f_n/f_p=-0.7$, a similar but less severe tension has now appeared, because the new CDMS-II-Si bounds are below the central values of all but the lowest energy modulated CoGeNT data, although the CoGeNT error bars extend well into the allowed region.

The recent CDMS-II-Si events appear in a similar situation of conflict with upper bounds if interpreted as due to dark matter WIMPs with spin-independent interactions, although the conflict is milder because the error bars are larger. In the isospin-conserving case, the central values of the CDMS-II-Si events are in the region excluded by the XENON100 bound, while the error bars of the two lowest energy CDMS-II-Si events extend into the allowed region. In the isospin-violating case $f_n/f_p=-0.7$, the CDMS-II-Si events correspond to a rate lower than the amplitude of the modulated rates in CoGeNT and DAMA. This makes it difficult to interpret the CoGeNT and DAMA annual modulations on one side and the CDMS-II-Si events  on the other as the modulated and unmodulated aspects of the same WIMP dark matter signal.

\bigskip

Making some assumptions about the halo, which apply when the function $\eta$ can be well approximated by the first two terms of its Taylor expansion in the Earth's orbital velocity $v_e$, Refs.~\cite{HerreroGarcia:2011aa, HerreroGarcia:2012fu} obtain more restrictive bounds on the modulation amplitude from upper bounds on the unmodulated rate.

\bigskip

After submitting this paper, we become aware of Ref.~\cite{Frandsen:2013cna}, which also gives an account of the compatibility of the new CDMS-II-Si signal with bounds from XENON100 and XENON10. There are the following differences between the two analyses: Ref.~\cite{Frandsen:2013cna} cuts $\mathcal{Q}_y$ to 0 in the XENON10 analysis while we do not, and uses a constant energy resolution $\sigma(E) = 0.3$ keV for CDMS-II-Si while we use the energy dependent resolution of the CDMS-II-Ge. Nevertheless, our results are very similar to those shown in Fig.~2 of Ref.~\cite{Frandsen:2013cna}, in which they use 2 keV bins as we do (the other figures in Ref.~\cite{Frandsen:2013cna} use 3 keV bins and are not directly comparable to ours).

\section*{Acknowledgments}

P.G. was supported in part by NSF grant PHY-1068111. E.D.N., G.G. and J.-H.H. were supported in part by DOE grant DE-FG02-13ER42022.
J.-H.H. was also partially supported by Spanish Consolider-Ingenio MultiDark (CSD2009-00064).

\end{document}